# WHAT CAUSES GEOMAGNETIC ACTIVITY DURING SUNSPOT MINIMUM?

**Kirov B.[1], Asenovski S.[1], Georgieva K.[1], Obridko V.N.[2]**
[1]*Space Research ant Technologies Institute – BAS, Sofia, Bulgaria*
[2]*IZMIRAN, Troitsk, Russia*

## ЧТО ОПРЕДЕЛЯЕТ ГЕОМАГНИТНУЮ АКТИВНОСТЬ В МИНИМУМЕ СОЛНЕЧНЫХ ПЯТЕН?

**Киров Б.[1], Асеновски С.[1], Георгиева К.[1], Обридко В.Н.[2]**
[1]*Институт Космических изследваний и технологий – БАН, София, Болгария*
[2]*ИЗМИРАН, Троицк, Россия*

*В работе показано, что средняя геомагнитная активность во время минимума солнечных пятен в последних 4 циклах последовательно убывает. Кроме того, установлено, что она не зависит от вариаций числа и/или параметров корональных выбросов массы и/или ударной волны, связанной с высокоскоростными потоками солнечного ветра. Показано, что у фонового солнечного ветра две компоненты: одна со скоростью до 450 км/с, другая – выше 490 км/с. Источник медленного ветра – гелиосферный токовый слой, а более быстрой компоненты – полярные корональные дыры. Средняя геомагнитная активность во время солнечного минимума определяется не только толщиной гелиосферного токового слоя, но и параметрами этих двух компонент солнечного ветра, которые изменяются от цикла к циклу.*

### 1. Introduction

Since the beginning of the geomagnetic measurements, the variations in the geomagnetic field have been related to solar activity. It is now known that big sporadic (non-recurrent) geomagnetic storms are caused by coronal mass ejections (CME). CMEs like sunspots are manifestations of the solar toroidal field and during sunspot maximum there is also a maximum in geomagnetic activity. Other sources of geomagnetic activity are the coronal holes – open unipolar magnetic field areas from which the high speed solar wind (HSS) emanates. Geomagnetic disturbances caused by HSS have maximum during the sunspots declining phase. These lead to two geomagnetic activity maxima in the 11-year sunspot cycle. In sunspot minimum, even during long periods without sunspots and without low-latitude coronal holes, geomagnetic disturbances are still observed.

Actually, geomagnetic activity can be divided into 3 components. The first one is the "floor", equal to $a_0$ coefficient which represents the geomagnetic activity in the absence of sunspots. It is practically determined by the activity in the cycle minimum and varies smoothly from cycle to cycle. The second component is the geomagnetic activity caused by sunspot-related solar activity which is described by the straight line $aa_T = b.R$ so that $aa_R = a_0 + aa_T$. The slope b of this line also changes cyclically. The third component $aa_P$ (the value above $aa_R$) is caused by HSS [1].

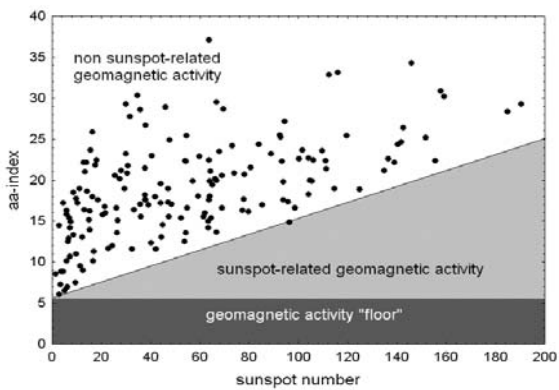

**Fig. 1.** Dependence of the geomagnetic activity on the sunspot number.

In this paper we look for the cause for the geomagnetic activity below $a_0$. For that we consider the geomagnetic activity in solar minima only.

## 2. Data

There are several catalogues for the solar wind parameters, covering the last four solar cycles. Using the experimental data presented in OMNIWeb Data Explorer (http://omniweb.gsfc.nasa.gov), we can study different parameters characterizing the solar wind state near the Earth for the conditions of solar minimum. For the purpose of our investigation we separate the data in four different sets – those which are measured one year before and one year after the last four solar minima. These data sets cover the periods: (20–21) min 1975–1977; (21–22) min 1985–1987; (22–23) min 1995–1997; (23–24) min 2007–2009. In addition, every data set was divided into three different types, covering periods with CMEs (Fig. 2), HSS (Fig. 3) and background solar wind. We categorize a CME when we have the following properties of the interplanetary space plasma:

– Proton temperature Tp < 0.5Tex, where Tex = 3(0.0106Vsw – 0.287) if Vsw < 500 km/s and Tex = (0.77Vsw – 265) if Vsw > 500 km/s (Vsw = Flow Speed).

– Magnetic field magnitude B > = 10 nT.

– Plasma Beta ≤ 0.8 for at least 5 hours.

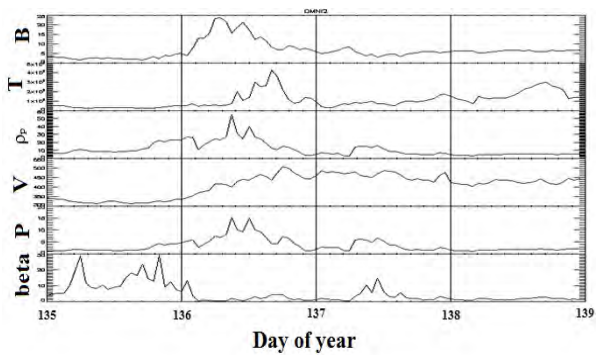 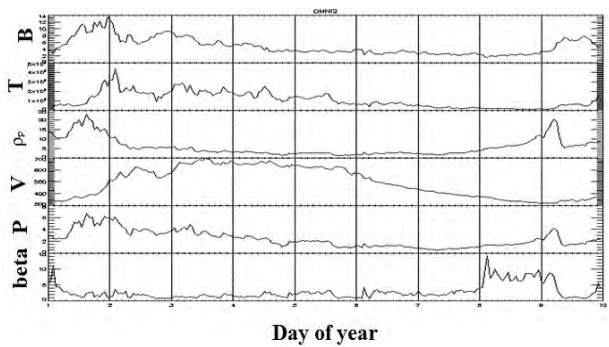

**Fig. 2.** Magnetic field B [nT], (first panel), Plasma Temperature T [K] (second panel), Proton Density N/cm$^3$ (third panel), Speed V km/s (forth panel), Flow pressure (fifth panel) and Total plasma beta (sixth panel) during a period of near Earth passing CME.

**Fig. 3.** Magnetic field B [nT], (first panel), Plasma Temperature T [K] (second panel), Proton Density N/cm$^3$ (third panel), Speed V km/s (forth panel), Flow pressure (fifth panel) and Total plasma beta (sixth panel) during a period of near Earth passing HSS.

The periods of HSS was determined by several catalogues [3–5].

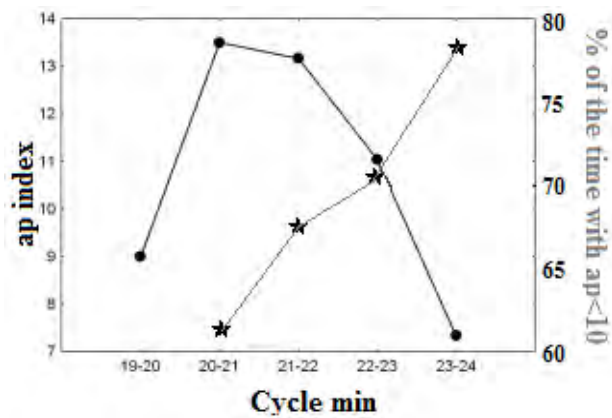 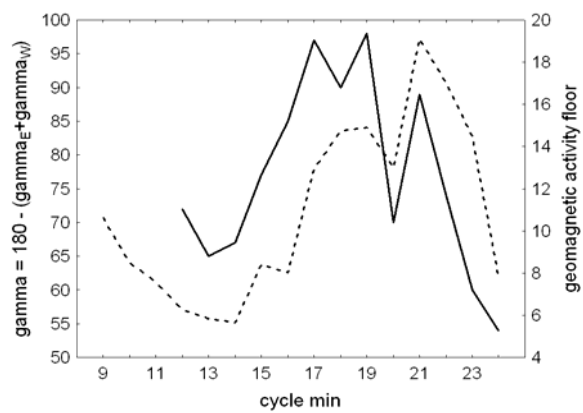

**Fig. 4.** Average Ap index in the last five solar minima and period of Ap < 10 index in the last four solar minima.

**Fig. 5.** Thickness of the helio-spheric current sheet (solid line) and geomagnetic activity floor (dashed).

### 3. Discussion

There is a declining trend of the average geomagnetic activity in the last four solar minima. This is caused by the increased period with Ap < 10 (Fig 4).

***What is the source of geomagnetic activity in sunspot minima?***

According to the recent theory, the geomagnetic activity even in solar minimum is caused by HSS and CME. On the one hand during the minima there is only a small number of CME events, and on the other hand most of the time the Earth is located inside the heliospheric current sheet and geomagnetic activity is lower than usual. The changes of geomagnetic activity are caused by:

❖ *Variations of the number and/or parameters of CME and/or HSS.* Our investigation shows that during the periods of minima, the time that the Earth is under the influence of CME is approximately 1,1% and the changes of the geomagnetic activity are not related to CME. The time that the Earth is under the influence of HSS is more than 50%, but there is no correlation between this time and geomagnetic activity (not shown), therefore we conclude that the number of HSS does not affect the geomagnetic activity significantly.

❖ *Variations of the thickness of the heliospheric current sheet.* Our investigation shows that the geomagnetic activity in the minimum is inversely proportional to the thickness [5] of the heliospheric current sheet (figure 5). The variations of the geomagnetic activity do not depend significantly on HSS and CME, but on the thickness of the heliospheric current sheet so we supposed that the main factor should be the background solar wind.

Using periods free of HSS and CME, we found out that the variations of background solar wind with speed less than 450 km/s influence the geomagnetic activity in a different way than the variation of background solar wind with speed greater than 490 km/s.

When the background solar wind is slower than 450 km/s, its average velocity during minima decreases, so does the average Ap index (Fig. 6). When the background solar wind is faster than 490 km/s, the increase of the average velocity leads to decrease of Ap index (Fig. 7).

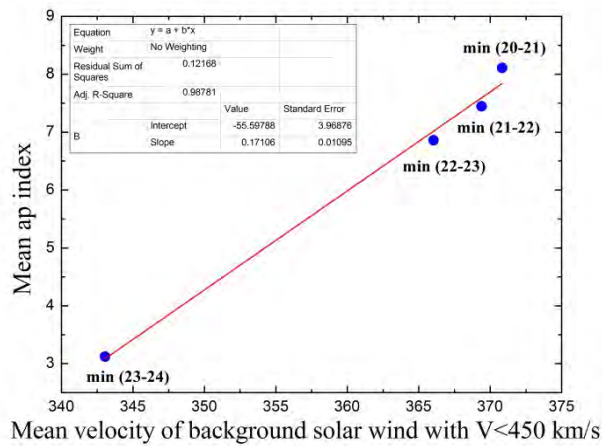 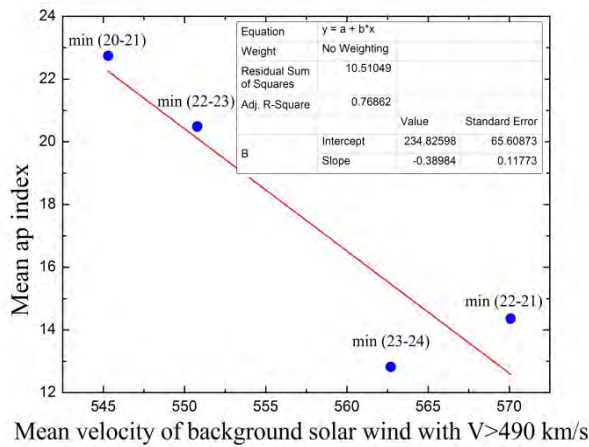

**Fig. 6.** Ap index in the last four solar minima in periods with background solar wind with V < 450 km/s.

**Fig. 7.** Ap index in the last four solar minima in periods with background solar wind with V > 490 km/s.

The background solar wind faster than 490 km/s and HSS have similar influence on the geomagnetic activity. Probably they have the same origin – polar coronal holes, while the slower background solar wind comes from heliospheric current sheet.

## 4. Conclusions

There are two factors that determine the average geomagnetic activity during sunspot minima:
– The thickness of the heliospheric current sheet;
– The parameters of the solar wind.
These averages are not related with CME and/or HSS.